\documentclass{jpsj3}

\title{Convergence Acceleration of Markov Chain Monte Carlo-based Gradient Descent by Deep Unfolding}

\author{Ryo Hagiwara\thanks{E-mail: hagiwara.r.ac@m.titech.ac.jp} and Satoshi Takabe}
\inst{Tokyo Institute of Technology, Ookayama, Tokyo, 152-8550, Japan}

\abst{
    {This study proposes} a trainable sampling-based solver for combinatorial optimization problems (COPs) {using} a deep-learning technique called deep unfolding. 
    The {proposed} solver is based on the Ohzeki method {that} combines Markov-chain Monte-Carlo (MCMC) and gradient descent, and its step sizes are trained by minimizing a loss function. 
    {In the training process, we propose a sampling-based gradient estimation that substitutes auto-differentiation with a variance estimation, thereby circumventing the failure of back propagation due to the non-differentiability of MCMC.}
    The numerical results for a few COPs {demonstrated} that the proposed solver {significantly accelerated} the convergence speed compared with the original Ohzeki method.
}

\usepackage{txfonts}
\usepackage{amsmath}
\usepackage{amssymb}
\usepackage{algorithm}
\usepackage{algpseudocode}
\usepackage{physics}
\usepackage{braket}
\usepackage{url}
\usepackage{color}
\usepackage{bm}
\usepackage{epstopdf}
\usepackage{graphicx}
\allowdisplaybreaks[3]

\begin{document}
\maketitle

{Combinatorial optimization problems (COPs) comprising discrete variables are considered hard to solve exactly in polynomial time, which relates to the well-known P vs. NP problem\cite{CTP}.}
Along with deterministic approximation algorithms, samplers such as Markov-chain Monte-Carlo (MCMC) have been applied to COPs. 
MCMC such as simulated annealing\cite{SA} for solving COP samples a candidate of solutions from the corresponding Boltzmann distribution. 
{However, the convergence time for obtaining reasonable approximate solutions is long. 
}

Recently, deep learning techniques have been applied {in several} research fields\cite{DL}. 
In particular, deep unfolding (DU)\cite{DU} has been used for developing new optimization-based algorithms in signal processing, such as wireless communications\cite{DUCS}, compressed sensing\cite{TISTA}, and image {processing}\cite{DUIM}.
{In contrast to other deep neural networks whose architecture is freely designed, the architecture of DU is based on an iterative optimization algorithm, wherein trainable parameters are embedded in the algorithm and optimized by deep learning techniques such as stochastic gradient descent and back propagation if the iterative processes are differentiable.}
{The convergence speed of the unfolded algorithm can be improved by learning trainable parameters,} whereas the number of parameters can be considerably smaller than a neural network\cite{Monga,Eldar}.
{Additionally}, for non-convex optimization problems, DU-based algorithms can improve the approximation performance compared to the original algorithm. {Hence}, various deterministic DU-based signal detectors have been proposed for solving COPs in wireless communications\cite{IncrementalTraining, THSDetector}.

Considering that a sampling-based COP solver is an iterative algorithm, it might be natural to apply DU to the solver in order to accelerate its convergence speed. 
However, {this} is not straightforward because, for example, most MCMC methods contain an acceptance--rejection process, which is a non-differentiable conditional branching process.  
{This can} lead to the failure of back propagation and auto-differentiation, {thereby} resulting in the hardness of optimizing trainable parameters of the unfolded algorithm. To the best of our knowledge, neither DU-based MCMC methods nor sampling-based COP solvers have been proposed to date. 

This {study proposes} a DU-based MCMC solver for COPs {by borrowing the structure of the Ohzeki method\cite{Ohzeki}, which is a combination of an MCMC sampler and a gradient descent}.
By introducing {and tuning} trainable step sizes, we attempt to accelerate the convergence speed.
Although the Ohzeki method includes a non-differentiable sampling part, {this study demonstrates} that the sampler can estimate the derivative instead of auto-differentiation, {enabling learning of trainable step sizes efficiently through back propagation.} 

{Here}, we focus on approximately solving a quadratic binary optimization with linear constraints defined by
\begin{equation}
    \begin{aligned}
    \min_{\bm{x} \in \{0,1\}^n} \quad  & f_0(\bm{x}) \\
    \textrm{s.t.} \quad & f_k(\bm{x}) = C_k \quad (k=1, \dots ,m),
    \end{aligned}\label{eq_COP}
\end{equation}
where $f_0$ is a quadratic or linear function and $f_k$ ($k=1,\dots,m$) is a linear function.
The COP (\ref{eq_COP}) is closely related to QUBO (quadratic unconstrained binary optimization)\cite{QUBO} and includes practical problems such as traveling salesman problems and graph partitioning problems.

The Ohzeki method\cite{Ohzeki} aims to find a solution {for} the COP (\ref{eq_COP}) efficiently using an MCMC sampler. 
{It} samples a solution from the Boltzmann distribution of an alternative cost function based on the penalty method defined as
\begin{equation}
    L(\bm{x}; \lambda) = f_0(\bm x)+\lambda \sum_{k=1}^m(f_k(\bm x)-C_k)^2
\end{equation}
where $\lambda (> 0)$ is a penalty parameter.
The auxiliary variables $\bm v\in\mathbb{R}^m$ for the penalty terms are introduced by the Hubbard--Stratonovich transformation\cite{Hubbard,Stratonovich}. 
Then, the gradient descent is employed to update $\bm v$ by
\begin{equation}
    v_k^{(t+1)} = v_k^{(t)} + \eta_{t}(C_k - \braket{f_k(\bm{x})}_{Q(\bm{v}^{(t)})}) \quad (t = 0, 1, \dots). \label{eq_OH} 
\end{equation}
In the above update rule, $\braket{\cdot}_{Q(\bm{v})}$ is the average over the Boltzmann distribution given by
\begin{equation} \label{OhzekiBoltmannDist}
    Q(\bm{v}^{(t)}) = \frac{1}{Z(\bm{v}^{(t)})}{\exp(-\beta f_0(\bm{x}) + \beta \sum_{k=1}^{m} v_{k}^{(t)} f_k(\bm{x}))},
\end{equation}
where $\beta$ is the inverse temperature and $Z(\bm{v}^{(t)})$ is the partition function. 
In Eq.~(\ref{eq_OH}), auxiliary variables $\bm v$ are updated to satisfy the linear constraints of the COP (\ref{eq_COP}). 
Different from the original constraints, the value of $f_k(\bm x)$ is replaced with the average $\braket{f_k(\bm{x})}_{Q(\bm{v})}$ over the auxiliary Boltzmann distribution (\ref{OhzekiBoltmannDist}). By taking the $\beta\to\infty$ limit, the optimal solution of the COP (\ref{eq_COP}) is obtained by sampling $\bm x$ from $Q(\bm{v}^\ast)$, where $\bm v^\ast$ is a convergent value of $\bm v$.  
The Ohzeki method is summarized as Alg.~\ref{alg_oh}.
{This method reduces the number of interactions compared with the penalty method that directly minimizes $L(\bm{x}; \lambda)$. }
It is suitable {particularly} for quantum annealing\cite{QA}, although any sampler is applicable to the Ohzeki method.
Note that the hardness of the original COP lies in the non-convexity of the function minimized by the gradient descent (\ref{eq_OH}). 
{Additionally}, the Ohzeki method repeats sampling from $Q(\bm{v}^{(t)})$ in each iteration, implying that the convergence speed of gradient descent largely affects the computational cost of the method.  
These indicate that the convergence speed and quality of solutions depend on the step sizes $\{\eta_{t}\}_{t=0}^{T-1}$.

\begin{algorithm}[t]
    \caption{Ohzeki method solving quadratic binary optimization with linear constraints}\label{alg_oh} 
    \begin{algorithmic}[1] 
    \State \textbf{Input:} $\beta$,$\lambda$,$f_0$, $\{f_k, C_k\}$, max iteration: $T$
    \State \textbf{Initialize:} $\bm{v}^{(0)} \in \mathbb{R}^m$
    \For{ $t = 0, 1, \dots ,T-1$}
    \State estimate $\{\braket{f_k(\bm{x})}_{Q(\bm{v}^{(t)})}\}$ by sampling from
    \State \quad $Q(\bm{v}^{(t)}) \propto \exp(-\beta f_0(\bm{x}) + \beta \sum_{k=1}^{m} v_{k}^{(t)} f_k(\bm{x}))$
    \State update $\bm{v}^{(t+1)}$ by
    \State \quad $v_k^{(t+1)} = v_k^{(t)} + \eta_{t} ( C_k - \braket{f_k(\bm{x})}_{Q(\bm{v^{(t)}})})$
    \EndFor\label{euclidendwhile}
    \State \textbf{return}  $\arg\min L(\bm{x}; \lambda)$ by sampling from $Q(\bm{v}^{(T)})$
    \end{algorithmic}
\end{algorithm}

DU enables {the learning of the internal parameters} of an iterative algorithm, leading to its convergence acceleration. Thus, it is expected that DU can tune the step sizes $\{\eta_t\}_{t=0}^{T-1}$ in the Ohzeki method. 
Here, the trainable method is named the deep-unfolded Ohzeki method (DUOM). 
Parameters of DUOM are trained by minimizing a loss function defined by $L \equiv \Braket{L(\bm{x}, \lambda)}_{Q(\bm{v}^{(T)})}$, where $\lambda$ is a hyperparameter.
The value is averaged over $Q(\bm{v}^{(T)})$ after $T$ iterations of DUOM with step sizes $\{\eta_t\}_{t=0}^{T-1}$.  
Minimizing $L$ {minimizes} the cost function $f_0$ and {satisfies} the constraints.
The training process is executed in an unsupervised manner {that is} different from {those used in} previous studies based on supervised learning~\cite{TISTA}.
This is because we assume a practical scenario where the optimal solution of the problem cannot be obtained in advance.

\begin{figure}[t]
    \centering
    \includegraphics[trim=0 0 0 0,width=0.7 \textwidth]{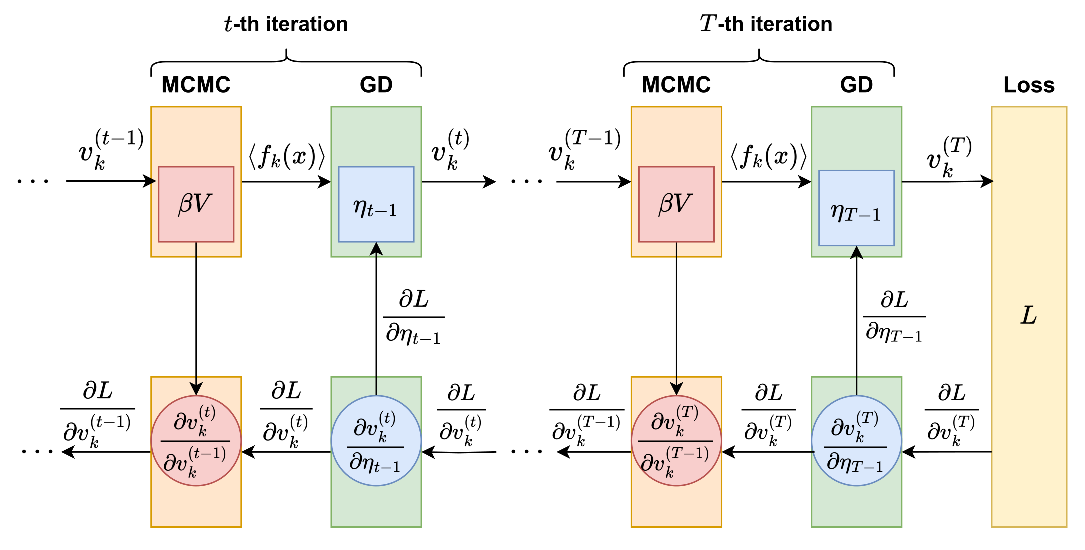}
    \caption{The architecture and training process of DUOM.
    The upper part represents a forward pass, whereas the lower part depicts a backward pass.
    DUOM {comprises} an MCMC sampler and gradient descent to update $\bm{v}$.
    In the forward pass, the expected value and variance $V$ of $\{f_k(\bm{x})\}$ are estimated by an MCMC sampler.
    Then, as the sampling-based gradient estimation, the variance is used to estimate the gradient in the backward pass, and trainable step sizes $\{\eta_t\}_{t=0}^{T-1}$ are updated by back propagation.
    Note that the process is executed simultaneously for $k=1,\dots,m$.
    }
    \label{DUOM_img}
\end{figure}

The training of unfolded algorithms is executed efficiently by back propagation if their internal processes are differential.
Back propagation calculates the partial derivatives of the trainable parameters using a backward pass, which is the backward process of a neural network, after evaluating the loss function using a forward pass \cite{BackProp}.  
In the case of DUOM, the chain rule in the backward pass is given by
\begin{align}
        \frac{\partial L}{\partial \eta_{t}} &= \sum_{k=1}^{m}\frac{\partial L}{\partial v_k^{(T)}}  \left(\prod_{u=t+1}^{T-1} \frac{\partial  v_k^{(u+1)}}{\partial{v_k^{(u)}}}\right)
        \frac{\partial v_k^{(t+1)}}{\partial \eta_t} , \label{eq_d1}\\
\frac{\partial v_k^{(u+1)}}{\partial v_k^{(u)}} &= 1 - \eta_{u} \frac{\partial \braket{f_k(\bm{x})}_{Q(\bm{v^{(u)}})}}{\partial v_k^{(u)}}, \label{eq_d2}
\end{align}
where $L$ is the loss function and Eq. (\ref{eq_d2}) is derived from Eq. (\ref{eq_OH}). 
However, Eq. (\ref{eq_d2}) contains the derivative of the average estimated by  a  non-differentiable sampler, leading to the failure of auto-differentiation.  
To circumvent {this}, we evaluate the derivative in advance{, expressed as}
\begin{align} 
\frac{\partial \braket{f_k(\bm{x})}}{\partial v_k}  &= \beta \left( \braket{f_k(\bm{x})} \Braket{\frac{\partial H }{\partial v_k}} -  \Braket{f_k(\bm{x}) \frac{\partial H}{\partial v_k}} \right), \nonumber\\
&=\beta (\braket{f_k^2(\bm x)}-\braket{f_k(\bm x)}^2), \label{eq_ex}
\end{align}
where $H= f_0(\bm{x}) - \sum_{k} v_{k}^{(u)} f_k(\bm{x})$ and $\braket{\cdot}=\braket{\cdot}_{Q(\bm{v^{(u)}})}$.
Therefore, substituting the variance of $f_k(\bm x)$ estimated by a sampler to the backward pass {executes} back propagation without auto-differentiation. 
We call the process sampling-based gradient estimation.
Compared with  a numerical derivative,
it is advantageous because the estimation of the variance is executed simultaneously with that of the average in the forward pass, {thereby} saving the call of the sampler and training parameters {rapidly}. 
The architecture and training process of DUOM is {shown} in Fig. \ref{DUOM_img}.

In the following, we show the results of numerical experiments to examine the performance of the proposed DUOM. In particular, we compare DUOM with the Ohzeki method with a fixed step size $\eta_t = \eta$.
Both methods employ the Metropolis--Hastings algorithm with a fixed $\beta$ as a sampler.

The first experiment is searching the $K$-minimum set among $N$ random numbers as a toy example.
The problem is selecting $K$ numbers out of $N$ uniformly distributed random numbers to minimize the sum of selected numbers{, defined as}
\begin{equation}
    \min_{\bm{x} \in \{0,1\}^n}  \sum_{i=1}^{N} h_i x_i \quad 
    \textrm{s.t.} \quad \sum_{i=1}^{N} x_i = K,
\end{equation}
where the variable $h_i$ is a random number independently following a uniform distribution $U(0,1)$.
The problem corresponds to $f_0(\bm x)=\sum_{i=1}^{N} h_i x_i$, $f_1(\bm x)=\sum_{i=1}^{N} x_i $, and $C_1= K$ with $m=1$ in the COP~(\ref{eq_COP}).

In the numerical simulation, we set $N=2000$, $K=50$, and $\beta=1000$. 
DUOM was implemented using pytorch 2.0.0\cite{pytorch} and learned by mini-batch training. 
As {this is} an unsupervised learning, the training data only {comprises} a random input $\bm{h} = (h_1, \dots , h_N)$.
In each parameter update, $50$ mini-batches of size $8$ are fed.
The parameters $\{\eta_t\}_{t=0}^{T-1}$ are updated by the Adam optimizer\cite{Adam} to minimize the loss function $L$ with $\lambda = 1$.
As a learning rate decay,  the learning rate was initialized by $5.0 \times 10^{-6}$ and reduced by a factor of $0.5$ for each update.
To avoid gradient vanishing, we used incremental training\cite{TISTA, IncrementalTraining}.
The number of iterations was set to $T = 20$, and {the initial values of the parameters were set to} $\eta_{t} = 1.0 \times 10^{-4}$ $(t = 0, \dots , T-1)$.

\begin{figure}[t]
    \centering
    \includegraphics[trim=0 0 0 0,width=0.7 \textwidth]{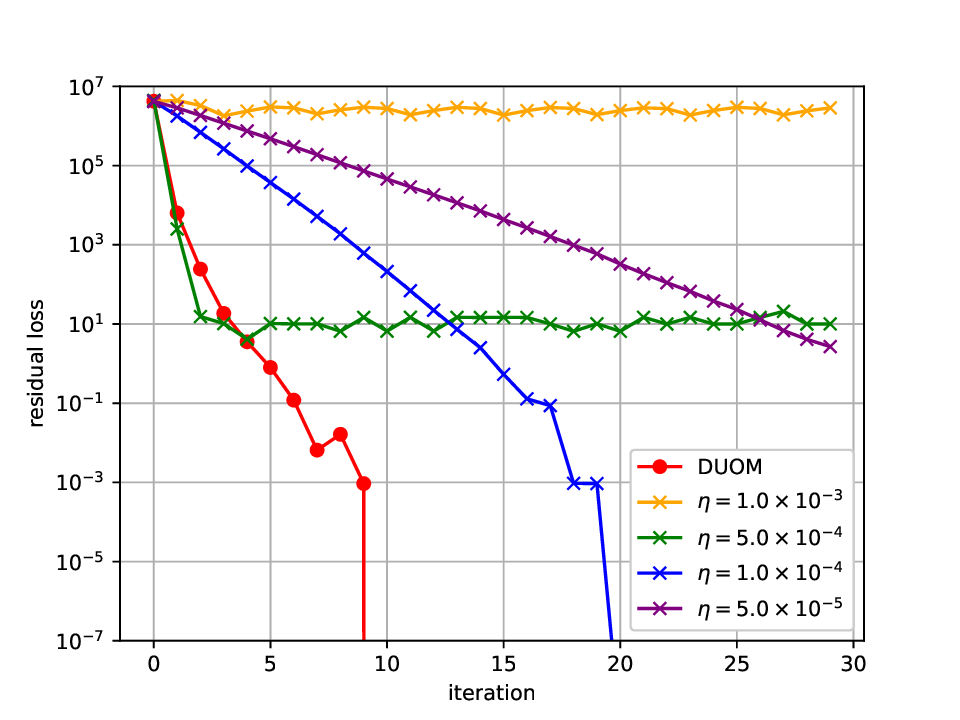}
    \caption{
            Iteration dependency of the residual loss by the Ohzeki method with a fixed step size $\eta_t = \eta$ and DUOM for the $K$-minimum set problem.  
        }
    \label{experiment1_cost_img}
\end{figure}

\begin{figure}[t]
    \centering
    \includegraphics[trim=0 0 0 0,width=0.7 \textwidth]{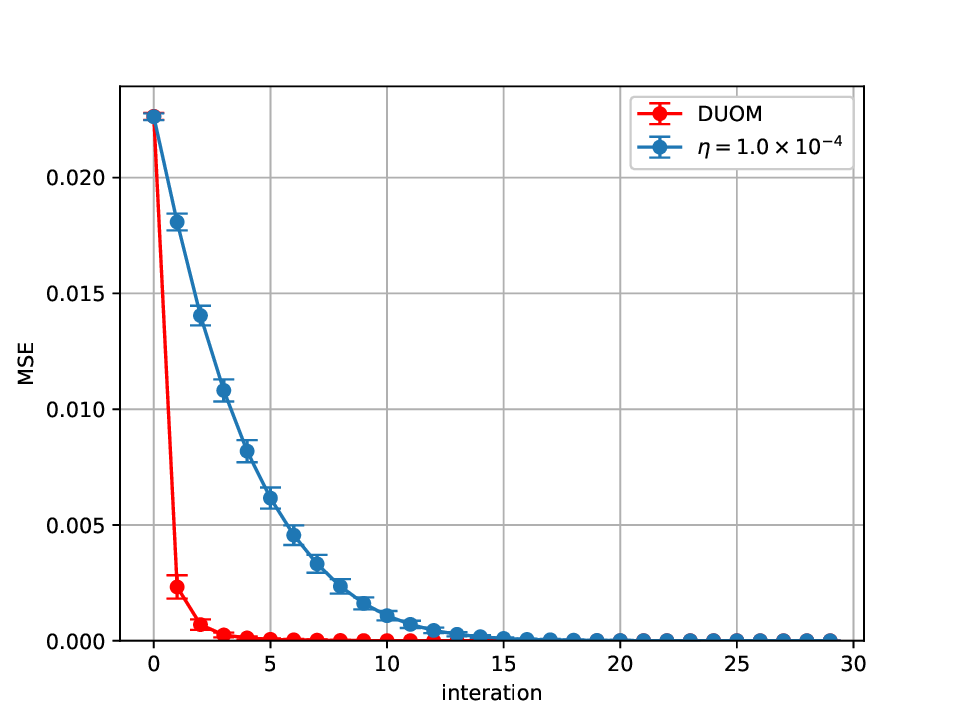}
    \caption{MSE performance of the Ohzeki method with $\eta = 1.0 \times 10^{-4}$  and DUOM as a function of the number of iterations.
        }
    \label{fig_experiment1}
\end{figure}

Figure \ref{experiment1_cost_img} shows the residual loss, which is the gap between the loss function $L$ of the output of a solver and of the corresponding optimal solution.
First, we found that the performance of the Ohzeki method depends on the value of a fixed step size.
{The performance was the best when $\eta = 1.0 \times 10^{-4}$; at other values, the method performed poorly.}
When the step size was too large, such as $\eta = 1.0 \times 10^{-3}$ and $5.0\times 10^{-4}$, the residual loss did not decrease quickly. 
{Conversely}, the convergence speed was slow when $\eta = 5.0 \times 10^{-5}$. 
This suggests that the step size should be tuned carefully to obtain considerable performance. 
In DUOM, the performance was reasonably improved, indicating successful learning of trainable step sizes. 
It was found that the residual loss dropped to $10^{-7}$ after $10$ iterations, whereas the best performance of the Ohzeki method with a fixed step size was {approximately} $10^{1}$ in $10$ iterations.
{Moreover}, the latter required at least $20$ iterations to achieve $10^{-7}$.
By learning step sizes, {improved performance was achieved with fewer iterations}.

Figure \ref{fig_experiment1} shows the mean squared error (MSE) $\|\bm{x} - \bm{x}^\ast \|_{2}^{2}/N$
between the output $\bm x$ and optimal solution $\bm x^\ast$ as a function of the number of iterations. 
{The MSE of DUOM decreased rapidly compared with that of the Ohzeki method with a fixed step size.}
For the $K$-minimum set problem, DUOM with trainable step sizes {improved} the convergence performance.
In contrast, the performance of the Ohzeki method with a fixed step size largely depends on the value $\eta$.

\begin{figure}[t]
    \centering
    \includegraphics[trim=0 0 0 0,width=0.7 \textwidth]{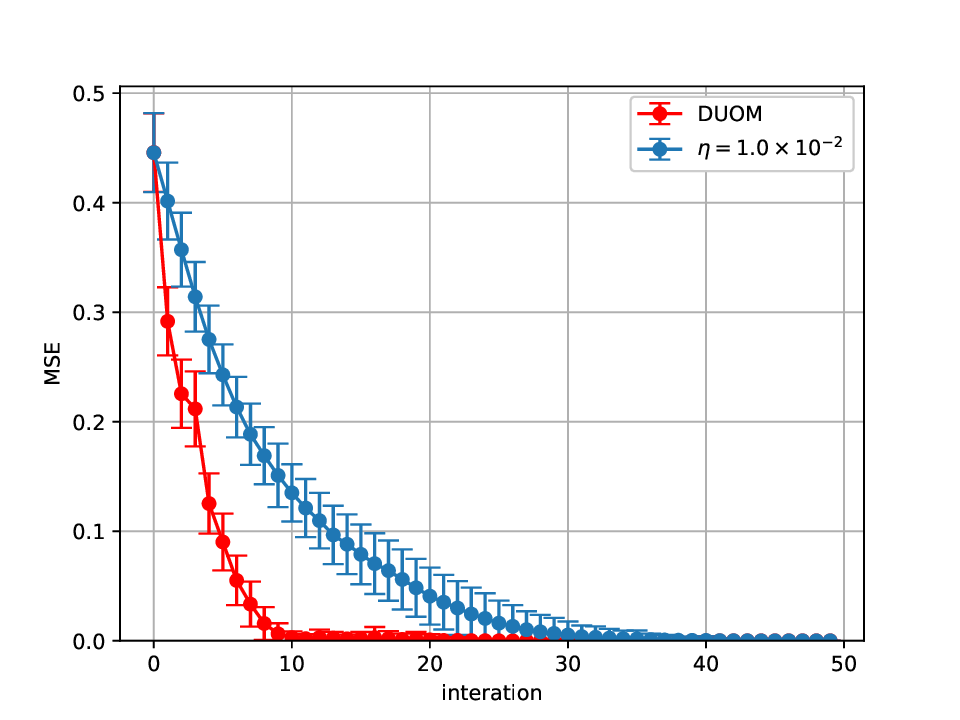}
    \caption{
        MSE as a function of the number of iterations for the image reconstruction problem.
        {The} Red symbols represent the results of DUOM, while {the} blue ones show the performance of the Ohzeki method with a fixed step size $\eta = 1.0 \times 10^{-2}$.
        }
    \label{fig_experiment2}
\end{figure}

Next, we conducted an experiment for image reconstruction by quadratic binary optimization with linear constraints~\cite{Ohzeki}.
Let $\bm{x}\in \{0,1\}^{N}$ be a vector representing a two-dimensional binary image of size $\sqrt N\times \sqrt N$.
We consider reconstructing $\bm{x}$ from the observation $\bm{y} = \bm{A} \bm{x}$, where $\bm{A}=(a_{kl})$ is an $M \times N$ ($M < N$) random matrix whose element follows the standard normal distribution {independently}. 
{Assuming that} non-zero elements connect to each other in the original image, the problem of reconstructing a binary image in the underdetermined system is formulated by
\begin{equation}
    \min_{\bm{x} \in \{0,1\}^N} \,\,  - \sum_{\braket{i,j}} x_i x_j \,\, 
       \textrm{s.t.} \quad  y_k = \sum_{l=1}^{N}a_{kl}x_l \,\, (k=1, \dots ,M), 
\end{equation}
where $\braket{i,j}$ represents a pair of indexes that are neighbors in the corresponding two-dimensional image. 
The problem corresponds to $f_0(\bm x)=-\sum_{\braket{ij}} x_i x_j$, $f_k(\bm x)= \sum_{l=1}^{N}a_{kl}x_l $, and $C_k= y_k$ ($k=1,\dots,M$) in the COP~(\ref{eq_COP}). 
Note that the cost function $f_0$ of the problem {comprises} quadratic forms different from the $K$-minimum set problem.

\begin{figure}[t]
    \centering
    \includegraphics[trim=0 0 0 0,width=0.7 \textwidth]{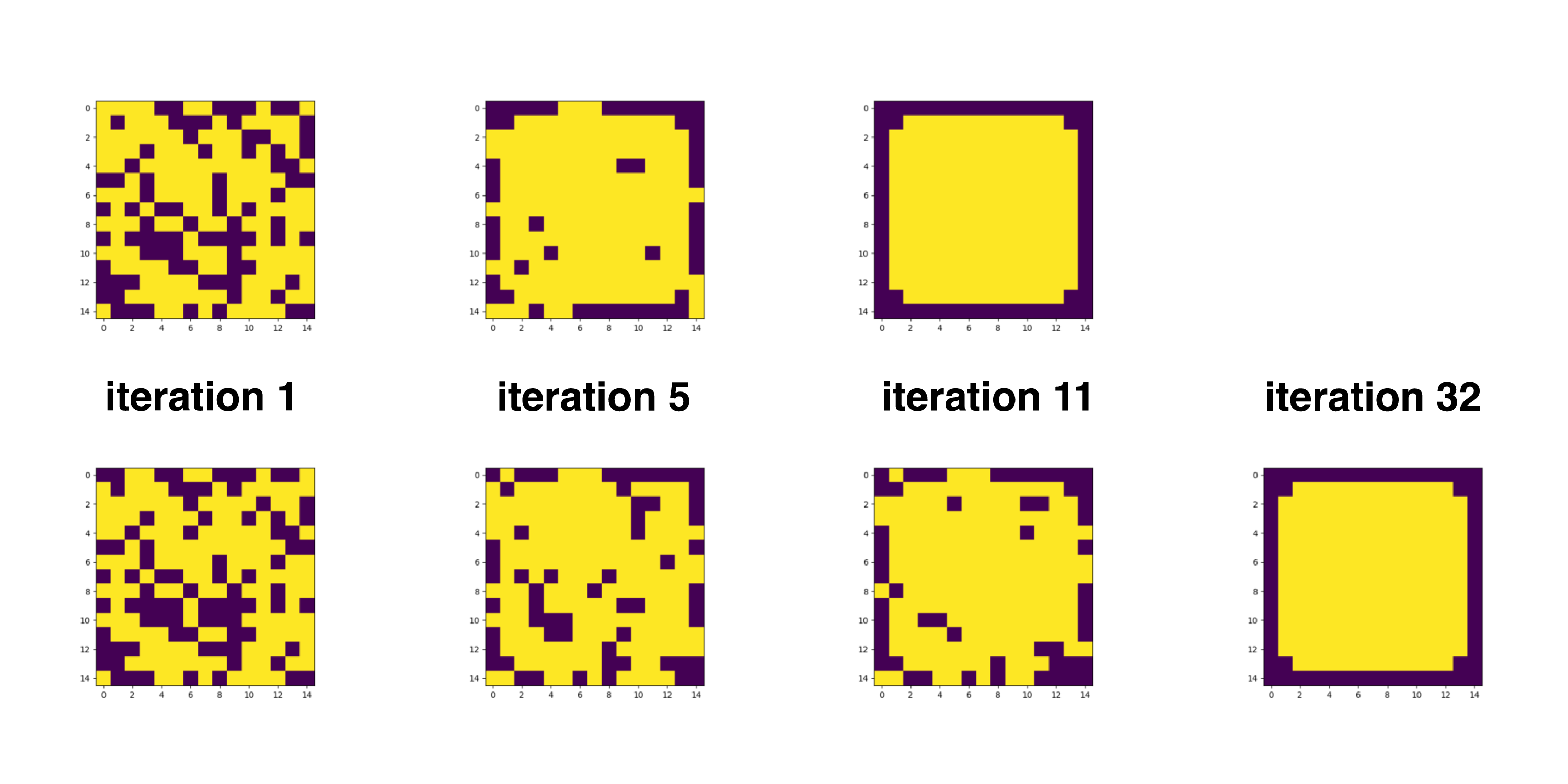}
    \caption{Examples of reconstructed images by DUOM (top) and the Ohzeki method with $\eta = 1.0 \times 10^{-2}$ (bottom).
    	 The size of the image is $15\times 15$, and the yellow and purple pixel represents a pixel of $x_i=1$ and $0$, respectively. 
    	 The reconstructed image of each iteration {was} chosen from samples to minimize the loss function $L$. 
    	 DUOM and the Ohzeki method with a fixed step size reconstructed the original image in $11$ and $32$ iterations, respectively.
        }
    \label{exsample_image_reconstruction}
\end{figure}

In the experiment, we set $N=15^2=225$, $M=135$, and $\beta=10$, {and} fixed the original image $\bm x^\ast$ as shown in Fig.~\ref{exsample_image_reconstruction} (top of ``iteration 11''). 
The fraction $M/N= 0.6 $ is lower than the statistical-mechanical threshold $M/N=0.633$~\cite{Tanaka}, above which the reconstruction can be executed by solving $\bm{y} = \bm{Ax}$.
For DUOM, training data {comprises} a random matrix $\bm{A}$ and $\bm{y}= \bm{A} \bm{x}^\ast$.
In each parameter update, $20$ mini-batches of size $8$ are fed, and the parameters $\{\eta_t\}_{t=0}^{T-1}$ were updated by an Adam optimizer\cite{Adam} to minimize $L$ with $\lambda = 1$.
We used incremental training and learning rate decay, where the learning rate was initialized by $5.0 \times 10^{-3}$ and reduced by a factor of $0.8$ for each update.
The number of iterations was set to $T = 40$, and the initial values of the trainable parameters were $\eta_{t} = 1.0 \times 10^{-2}$ $(t = 0, \dots , T-1)$.

Figure \ref{fig_experiment2} shows the MSE between the original image and reconstructed ones by DUOM and the Ohzeki method with a fixed step size. 
The value of a fixed step size was set to $\eta = 1.0 \times 10^{-2}$, which is optimized by a grid search. 
{The results showed} that the convergence speed of DUOM {was} significantly accelerated by learning {the} step sizes. 
In particular, the MSE of DUOM became zero within 30 steps, whereas the MSE of the Ohzeki method with a fixed step size decreased {gradually}. 

Figure \ref{exsample_image_reconstruction} shows an example of reconstructed images by DUOM and the Ohzeki method with a fixed step size $\eta = 1.0 \times 10^{-2}$. 
Each output $\bm{x}^{(t)}$ is displayed as a two-dimensional binary image.  
In the experiment, DUOM converged to the original image in $11$ iterations, whereas the Ohzeki method reconstructed the original image in $32$ iterations.
{The results showed} that DU-based parameter learning successfully {improved} the convergence performance of the Ohzeki method even when the cost function $f_0$ {contained} quadratic terms.

{In conclusion, this study proposed} a trainable MCMC-based solver {named DUOM} for a binary quadratic optimization with linear constraints. 
The solver was inspired by the Ohzeki method, whose performance depends on step sizes in gradient descent.  
By combining the Ohzeki method with DU, DUOM can learn its step sizes efficiently by deep learning techniques. 
Because the Ohzeki method contains a non-differential process related to a sampler, we proposed an alternative sampling-based gradient estimation by estimating the variance. 
This enabled us to use back propagation efficiently in the training process.  
We conducted numerical experiments for the $K$-minimum set and binary image reconstruction. 
The results {showed} that DUOM {has} significantly {improved} the performance compared {to that of} the original Ohzeki method. 
In particular, DUOM {accelerated} the convergence speed, {thereby reducing} the number of sampler calls and the resulting computational cost. 

It is known that the Ohzeki method is suitable for quantum annealing because it reduces the number of interactions and problem size  in practice.~\cite{Ohzeki}.
Although a classical MCMC is employed in this {study}, applying the proposed DUOM to quantum annealing is an interesting {aspect for} future work.
Another direction is the application of DUOM to other NP-hard problems such as a traveling salesman problem and graph partitioning.

\acknowledgement{ST {is grateful to} S. Arai for fruitful discussions.
This work was partly supported by JSPS Grant-in-Aid for Scientific Research Grant Numbers 22H00514 and 22K17964.
}

\end{document}